\documentclass[10pt,conference]{IEEEtran}
\IEEEoverridecommandlockouts
\usepackage{cite}
\usepackage{amsmath,amssymb,amsfonts}
\usepackage{algorithmic}
\usepackage{graphicx}
\usepackage{textcomp}
\usepackage{xcolor}
\def\BibTeX{{\rm B\kern-.05em{\sc i\kern-.025em b}\kern-.08em
    T\kern-.1667em\lower.7ex\hbox{E}\kern-.125emX}}

\usepackage{caption}
\usepackage{subcaption}
\usepackage[framemethod=tikz]{mdframed}
\usepackage{url}
\usepackage{booktabs}
\usepackage{makecell}
\usepackage{tabularx}
\newcolumntype{Y}{>{\centering\arraybackslash}X}
\usepackage{mathabx}

\renewcommand{\geq}{\vargeq}

\usepackage{stfloats} 
\usepackage[normalem]{ulem}
\begin{document}

\title{Performance Analysis of IEEE 802.11bn with Coordinated TDMA on Real-Time Applications\\

\vspace{-10pt}
\author{\IEEEauthorblockN{Seungmin Lee\IEEEauthorrefmark{1},
Changmin Lee\IEEEauthorrefmark{1},
Si-Chan Noh\IEEEauthorrefmark{2}, and
Joonsoo Lee\IEEEauthorrefmark{2}}
\thanks{This work was supported by the Technology Innovation Program(20023975, Ultra LowPower Long Range WiFi SoC Research Development for IoT Devices) funded By the Ministry of Trade Industry \& Energy(MOTIE, Korea)}
\IEEEauthorblockA{\IEEEauthorrefmark{1}Newratek, Seoul, South Korea}
\IEEEauthorblockA{\IEEEauthorrefmark{2}Newracom, Irvine, CA, USA}
\IEEEauthorblockA{Emails: \IEEEauthorrefmark{1}\{sm.lee, cm.lee\}@newratek.com; \IEEEauthorrefmark{2}\{sc.noh, js.lee\}@newracom.com}
}
}

\maketitle

\begin{abstract}
Wi-Fi plays a crucial role in connecting electronic devices and providing communication services in everyday life. Recently, there has been a growing demand for services that require low-latency communication, such as real-time applications. The latest amendments to Wi-Fi, IEEE 802.11bn, are being developed to address these demands with technologies such as the multiple access point coordination (MAPC). In this paper, we demonstrate that coordinated TDMA (Co-TDMA), one of the MAPC techniques, effectively reduces the latency of transmitting time-sensitive traffic. In particular, we focus on worst-case latency and jitter, which are key metrics for evaluating the performance of real-time applications. We first introduce a Co-TDMA scheduling strategy. We then investigate how this scheduling strategy impacts latency under varying levels of network congestion and traffic volume characteristics. Finally, we validate our findings through system-level simulations. Our simulation results demonstrate that Co-TDMA effectively mitigates jitter and worst-case latency for low-latency traffic, with the latter exhibiting an improvement of approximately 24\%.
\end{abstract}


\vspace{-15pt}

\section{Introduction}
\label{Intro}

In daily life, individuals rely extensively on communication among various devices to access numerous services. The communication technology known as IEEE 802.11, commonly referred to as Wi-Fi, plays a pivotal role in facilitating these services~\cite{IEEE802.11-2020}. Recently, new demands have emerged for real-time applications, including augmented/virtual reality (AR/VR) applications, real-time mobile gaming, and industrial internet of things (IIoT). Real-time applications require low latency and minimal packet loss to deliver satisfactory user experiences~\cite{Meng2019}. To satisfy these emerging requirements, new IEEE 802.11 standards, such as IEEE 802.11be and IEEE 802.11bn, are under active development~\cite{Garcia2021, Reshef2022, Giordano2024}.

Fundamentally, Wi-Fi employs a contention-based mechanism for devices to obtain transmission opportunities (TXOPs). Although contention-based transmission simplifies selecting transmitting devices, excessive contention can degrade network efficiency. Hence, Wi-Fi advancements have aimed to alleviate contention and enhance efficient network resource utilization. For instance, IEEE 802.11ac amendment added a mechanism enabling TXOP sharing across different access categories (ACs) within a station (STA). In IEEE 802.11ax amendment, the role of access points (APs) was expanded, introducing several functionalities to efficiently manage network resources within a basic service set (BSS). Among these functionalities, uplink multi-user (UL MU) operation enables an AP to coordinate simultaneous transmissions from multiple non-AP STAs, reducing channel access contention among STAs. IEEE 802.11be amendment introduced triggered TXOP sharing, allowing AP to delegate TXOP to a specific non-AP STA, enabling them to transmit urgent or large volumes of data without being constrained by AC restrictions.

Building upon this, IEEE 802.11bn amendment aims to accommodate requirements of emerging services by introducing ultra-high reliability features, incorporating enhancements to throughput, reduced latency, and decreased medium access control (MAC) protocol data unit (MPDU) loss~\cite{Giordano2024, PAR, IEEE802.11bn}. One of the key schemes is multiple access point coordination (MAPC), a mechanism that enables multiple APs to collaborate in data transmission. In MAPC, APs perform discovery and negotiation processes to establish coordination schemes. These processes can be carried out via either wireless or wired communication channels~\cite{Giordano2024}. The MAPC framework enables functionalities such as coordinated TDMA (Co-TDMA), coordinated beamforming (Co-BF), coordinated spatial reuse (Co-SR), and coordinated restricted target wake time (Co-RTWT)~\cite{IEEE802.11bn}. In particular, Co-TDMA facilitates TXOP sharing among APs, enhancing the efficient utilization of the limited network time resource, thereby improving the latency of real-time applications~\cite{Ver2019, Noh2023, Noh2024, Val2024}.

\begin{figure*}[t]
    \centering
    \includegraphics[width=.85\linewidth]{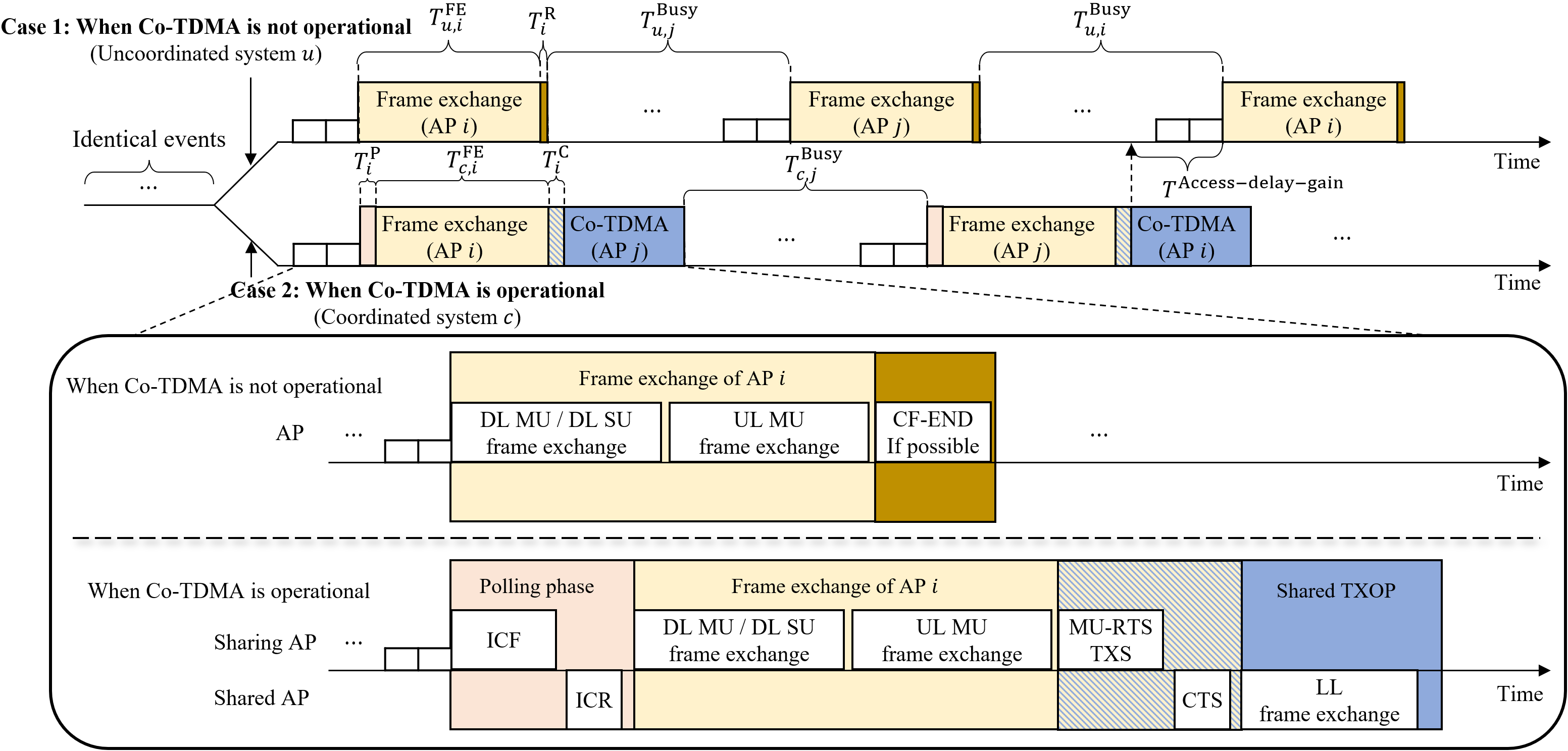}
    \caption{Frame exchange examples for coordinated system and uncoordinated system.}
    \label{fig:CoTDMA_schme}
    \vspace{-10pt}
\end{figure*}

In this paper, we evaluate whether Co-TDMA can effectively contribute to reducing latency. Additionally, we investigate critical factors for effective Co-TDMA utilization and verify it through simulations applying a real-time application traffic model within an enterprise environment comprising multiple overlapping basic service sets (OBSS).
We anticipate that Co-TDMA can effectively reduce the latency of time-sensitive traffic generated by real-time applications.
The key contributions of this paper are summarized as follows:

\begin{itemize}
    \item We demonstrate that an appropriate scheduling strategy with Co-TDMA can improve latency while effectively maintaining network throughput.
    \item We evaluate latency enhancements provided by Co-TDMA relative to varying levels of network congestion.
    \item We observe how traffic characteristics influence the trend of latency. Each real-time application exhibits diverse traffic characteristics concerning packet size, frequency, and other factors.
\end{itemize}

\section{Related Work}
The MAPC framework is an active area of research, and here we introduce several studies focusing on Co-TDMA, one of its key sub-fields.
In~\cite{Nunez2022}, the authors investigated Co-TDMA/SR, demonstrating improvements in throughput and reduction of transmission duration. Performance validation was conducted using simulations that considered varying network topologies.
In~\cite{Val2024}, the authors presented simulation results for Co-TDMA scenarios within residential environments, focusing on throughput and latency metrics. They assumed that only traffic belonging to the same AC is transmitted within a shared TXOP. Under these conditions, the results illustrated how Co-TDMA performance varies depending on the Co-TDMA strategies and network characteristics.
In~\cite{Zhang2025}, the authors proposed an optimal transportation channel access (OTCA) mechanism for a Co-TDMA-based network. Subordinate APs provide Co-TDMA-related information to a centralized AP via wired links to coordinate resource utilization. 
They demonstrated improved throughput and latency by conducting simulations based on OTCA.

\section{Co-TDMA}
\label{co-tdma scheme}

Time-sensitive traffic generated by real-time applications requires low-latency (LL) communication. Thus, we refer to this traffic as LL traffic. Reducing the worst-case latency and jitter of LL traffic is essential to supporting real-time applications; therefore, minimizing the latency between non-AP STA and AP is crucial to improving the latency experience in real-time applications over Wi-Fi~\cite{Meng2019}. Thus, we consider end-to-end (E2E) latency as a major performance metric. E2E latency consists of queuing delay, channel access delay, retransmission delay, and transmission delay, representing the time that elapses from the moment a packet enters the queue for transmission until it is successfully received~\cite{Naik2020}. Meanwhile, Co-TDMA has been proposed to improve worst-case latency while maintaining throughput fairness~\cite{Ver2019}. In the following sections, we investigate the effectiveness of Co-TDMA in reducing both worst-case and jitter of E2E latency.

Co-TDMA is an operation that enables an AP to share a time portion of an obtained TXOP with one of more APs.
It allows TXOP sharing among multiple APs; however, in this paper, we consider a scenario where only a pair of APs can perform Co-TDMA. Under this assumption, we can observe the impact of Co-TDMA on OBSS when coordinated BSSs that support this mechanism coexist with BSSs, including legacy ones.

\begin{figure*}[t]
\begin{align}
    {T}^{\text{Access-delay-gain}}_{i} &= 
    \overbrace{{T}^{\text{Overhead}}_{u} \! + \! \sum_{k=i,j} {T}^{\text{FE}}_{u,k} \! + \! {T}^{\text{Busy}}_{u,j} \! + \! {T}^{\text{Busy}}_{u,i}}^{\textstyle \textstyle {T}^{\text{Access-interval}}_{u,i}}  \! - \! \overbrace{\left( {T}^{\text{Overhead}}_{c} \! + \! \sum_{k=i,j} {T}^{\text{FE}}_{c,k} \! + \! {T}^{\text{Busy}}_{c,j} \! + \! {T}^{\text{Co-TDMA}} \right)}^{\textstyle \textstyle {T}^{\text{Access-interval}}_{c,i}}   \label{equation:1}\\
    &\geq {T}^{\text{Overhead}}_{u} \! + \! {T}^{\text{Busy}}_{u,j} \! + \! {T}^{\text{Busy}}_{u,i} \! - \! \left( {T}^{\text{Overhead}}_{c} \! + \! {T}^{\text{Busy}}_{c,j} \! + \! {T}^{\text{Co-TDMA}} \right), \quad\quad\quad\quad\!\; \text{if ${T}^{\text{FE}}_{u,k} = {T}^{\text{FE}}_{c,k}, \; \forall{k},$}  \label{equation:2} \\
    &\approx {T}^{\text{Busy}}_{u,i} \! - \! {T}^{\text{Co-TDMA}}, \quad\quad\quad\quad\quad\quad\quad\quad\quad\quad\quad\quad\quad\quad\quad\quad\quad\quad\quad\;\;\;\;\;\;\;\!\!\! \text{if ${T}^{\text{Busy}}_{s,k}$ is large enough, \;$\forall{s,k}.$} \label{equation:3}
\end{align}
\hrule
\vspace{-16pt}
\end{figure*}

An AP that shares a portion of its obtained TXOP with another AP is referred to as the sharing AP, while an AP that is allocated the TXOP is referred to as the shared AP. Co-TDMA procedure begins with the exchange of an initial control frame (ICF) and an initial control response (ICR) within a TXOP. This process is referred to as the Co-TDMA polling phase, during which the sharing AP identifies potential candidate APs and obtains the necessary information for 
Co-TDMA scheduling. After the polling phase, the sharing AP uses a portion of its TXOP for frame exchange with its own associated STAs and then allocates the TXOP by transmitting an initiating frame, referred to as the multi-user request to send TXOP sharing (MU-RTS TXS) trigger frame
.
Upon reception, the shared AP responds with a CTS frame, thereby confirming the initiation of the sharing TXOP. The shared AP then starts frame exchanges during the shared TXOP duration~\cite{IEEE802.11bn}. This whole procedure is illustrated in Fig. \ref{fig:CoTDMA_schme}. It shows frame exchange examples of a coordinated system supporting Co-TDMA and an uncoordinated system that does not support MAPC functionality.

\subsection{Scheduling strategy}
The latency experienced by LL traffic is dependent on the timing and frequency of Co-TDMA operation. In this paper, we introduce a Co-TDMA scheduling method aimed at improving latency experienced by LL traffic.

Firstly, we describe several assumptions required for Co-TDMA operation.
\begin{itemize}
    \item A sharing AP can perform Co-TDMA operations during TXOPs assigned to any AC, except its TXOP limit is zero~\cite{IEEE802.11bn}.
    \item Sharing AP and candidate shared APs become aware of each other’s presence through management frame exchange over wireless channels or via wired backhaul links~\cite{IEEE802.11bn, Giordano2024}.
    \item The information required among APs with MAPC can be obtained through the ICF-ICR frame exchange or the backhaul link~\cite{IEEE802.11bn, Giordano2024}.
    \item Candidate shared APs can identify the presence of LL traffic generated by its associated STAs, by using existing mechanisms such as stream classification service (SCS) or traffic specification (TSPEC). Additional mechanisms for identifying the LL traffic status on the STA side are currently under discussion in~\cite{IEEE802.11bn}.
\end{itemize}

In general, the primary purpose of acquiring a TXOP is to earn the opportunity to transmit frames in its own queue. Therefore, to guarantee the transmission rights of the AP that wins the channel access contention, downlink transmission is given the highest priority in all TXOPs. The next priority is based on the closeness of the relationship with the AP. Therefore, the sequence of actions that the sharing AP can take within a TXOP is summarized as follows:
\begin{enumerate}
    \item Attempt downlink multi-user (DL MU) or downlink single-user (DL SU) transmissions if DL traffic exists.
    \item If DL is not feasible, attempt uplink multi-user (UL MU) transmissions for only LL traffic if the identified UL LL traffic size is greater than zero.
    \item If the DL and UL procedures have been done, attempt Co-TDMA operation if a candidate shared AP responded during polling phase.
    \item If no transmission is possible, send a contention-free end (CF-End) frame to truncate the TXOP.
\end{enumerate}

The shared TXOP duration is defined as the sum of the time required for the shared AP to transmit both DL and UL of LL traffic.
The duration is computed by the sharing AP based on information provided by candidate shared APs via the backhaul link or during the polling phase. Candidate shared APs can obtain this information through mechanisms such as TSPEC or SCS.
The shared AP can transmit LL traffic regardless of the AC of the shared TXOP.

\subsection{Latency analysis}
E2E latency is strongly related to channel access delay. Specifically, the E2E latency experienced by a frame is closely tied to the frequency and timeliness of channel access opportunities following the frame’s arrival at the sender. Consequently, increasing channel access opportunities within a given period reduces E2E latency~\cite{Naik2020}. However, such behavior must not significantly elevate the network collision rate. Co-TDMA assesses network idleness through a control frame exchange before sharing. That is, Co-TDMA can provide more frequent channel access opportunities without substantially increasing collision rates. 
Thus, it can contribute to improved E2E latency by providing more frequent channel access opportunities for a period. Here, under our scheduling strategy, we analyze how much channel access delay can be reduced in a coordinated system compared to an uncoordinated system when an AP gains a channel access opportunity by Co-TDMA.

We assume that randomness is identical in both systems, as depicted in Fig. \ref{fig:CoTDMA_schme}.
That is, all event sequences prior to the occurrence of a Co-TDMA event are identical. 
In addition, we assume that APs are positioned within the coverage areas of each other's power detection (PD) and energy detection (ED) boundaries.
Finally, we consider a homogeneous scenario where the traffic distribution is identical across all BSSs.

Let APs capable of Co-TDMA be denoted as AP $k$ ($k=\{i,j\}$). 
\( T^{\text{Access-interval}}_{u,i} \) and \( T^{\text{Access-interval}}_{c,i} \) represent the channel access delays experienced by AP \( i \) from the start of a successful first channel access to the start of a successful second channel access, in the uncoordinated system \( u \) and the coordinated system \( c \), respectively.

Then, $T^{\text{Access-interval}}_{u,i} - T^{\text{Access-interval}}_{c,i}$ is a gain $T^{\text{Access-delay-gain}}_{i}$, meaning that the second channel access time of AP \(i\) in the coordinated system can be advanced by \(T^{\text{Access-delay-gain}}_{i} \) compared to the uncoordinated system. 
\( T^{\text{Access-interval}}_{u,i} \) can be decomposed into the time components \{$T^{\text{FE}}_{u,k}$, ${T^{\text{Overhead}}_u}$, $T^{\text{Busy}}_{u,k}|k=\{i,j\}$\}. \( T^{\text{Access-interval}}_{c,i} \) can be decomposed as \(\{ T^{\text{FE}}_{c,k}, T^{\text{Overhead}}_c, T^{\text{Busy}}_{c,j}, T^{\text{Co-TDMA}}|k=\{i,j\}\}\).

\( T^{\text{FE}}_{s,k} \) represents the frame exchange duration of AP \(k\) in the system \(s\) ($s=\{u,c\}$). \(T^{\text{Busy}}_{s,k} \) represents the busy duration experienced by AP \( k \) in system \( s \) after the TXOP of AP \( i \) or AP \( j \) has ended. 
$T^{\text{Overhead}}_{s}$ indicates the overhead duration included in $T^{\text{Access-interval}}_{s,i}$ in system $s$. Specifically, $T^{\text{Overhead}}_u$ is defined as $\sum_{k=i,j}{T^{\text{R}}_k}$, where $T^{\text{R}}_k$ represents the remaining TXOP duration of AP $k$ in system $u$, and $T^{\text{R}}_k$ is bounded by the duration of the CF-END frame. Meanwhile, $T^{\text{Overhead}}_c$ is defined as $\sum_{k=i,j}{({T^{\text{P}}_k + T^{\text{C}}_k})}$, where ${T^{\text{P}}_k}$ and ${T^{\text{C}}_k}$ correspond to the durations required for the polling phase and for exchanging control frames in system $c$, respectively.

During \( T^{\text{Access-interval}}_{c,i} \), at most one Co-TDMA operation can occur since only a pair of APs (\( i \) and \( j \)) can perform Co-TDMA.
We refer to the duration as \( T^{\text{Co-TDMA}} \).
Transmissions of all STAs and APs, except AP \( k \), are reflected in \( {T}^{\text{Busy}}_{s,k} \). Then, \( T^{\text{Access-delay-gain}}_{i} \) can be expressed as shown in (1).

If the amount of traffic that can be transmitted within the TXOP of AP \( i \) is greater than or equal to the TXOP limit, then \( {T}^{\text{FE}}_{u,i} = {T}^{\text{FE}}_{c,i} \), as AP \( i \) does not initiate Co-TDMA due to insufficient remaining TXOP. Otherwise, \( {T}^{\text{FE}}_{u,i} = {T}^{\text{FE}}_{c,i} \) holds since both systems have identical buffered traffic before ${T}^{\text{FE}}_{s,i}$. 
And, \( {T}^{\text{FE}}_{u,j} \geq {T}^{\text{FE}}_{c,j} \) always holds, since AP $j$ in system $c$ has already transmitted traffic through the preceding shared TXOP. 
When \( {T}^{\text{FE}}_{u,j} = {T}^{\text{FE}}_{c,j} \), the channel access delay gain is always smaller than that presented in~(1), and is expressed as in~(2).

The primary objective of Co-TDMA is to improve latency experience in congested networks~\cite{Giordano2024}. 
As network congestion increases, the term \( {T}^{\text{Overhead}}_{u} - {T}^{\text{Overhead}}_{c} \) becomes negligible compared to ${T}^{\text{Busy}}_{s,k}$. 
Furthermore, in both systems, channel access parameters of other APs and STAs before AP \( i \)’s first TXOP remain identical, and these stay frozen during frame exchanges. 
Particularly, the contention environment is nearly identical at the end of the first TXOP in systems \(c\) and \(u\), as the Co-TDMA duration is typically short due to the limited remaining TXOP duration under our scheduling strategy.
Therefore, we assume that when network contention is sufficiently high, \( {T}^{\text{Busy}}_{u,j} \approx {T}^{\text{Busy}}_{c,j} \) holds.
Consequently, (\ref{equation:2}) can be approximated as \( {T}^{\text{Busy}}_{u,i} - {T}^{\text{Co-TDMA}} \), as given in (\ref{equation:3}). 

This implies that as network congestion and contention intensify, the channel access delay gain achieved by Co-TDMA increases, potentially improving in E2E latency in the coordinated system. However, we note that the rate of increase in the channel access delay gain diminishes, as 
${T}^{\text{Busy}}_{s,k}$ grows logarithmically with network congestion~\cite{Yang2003}. We demonstrate this through comprehensive simulations.

\begin{figure} [t]
    \centering
    \includegraphics[width=.88\linewidth]{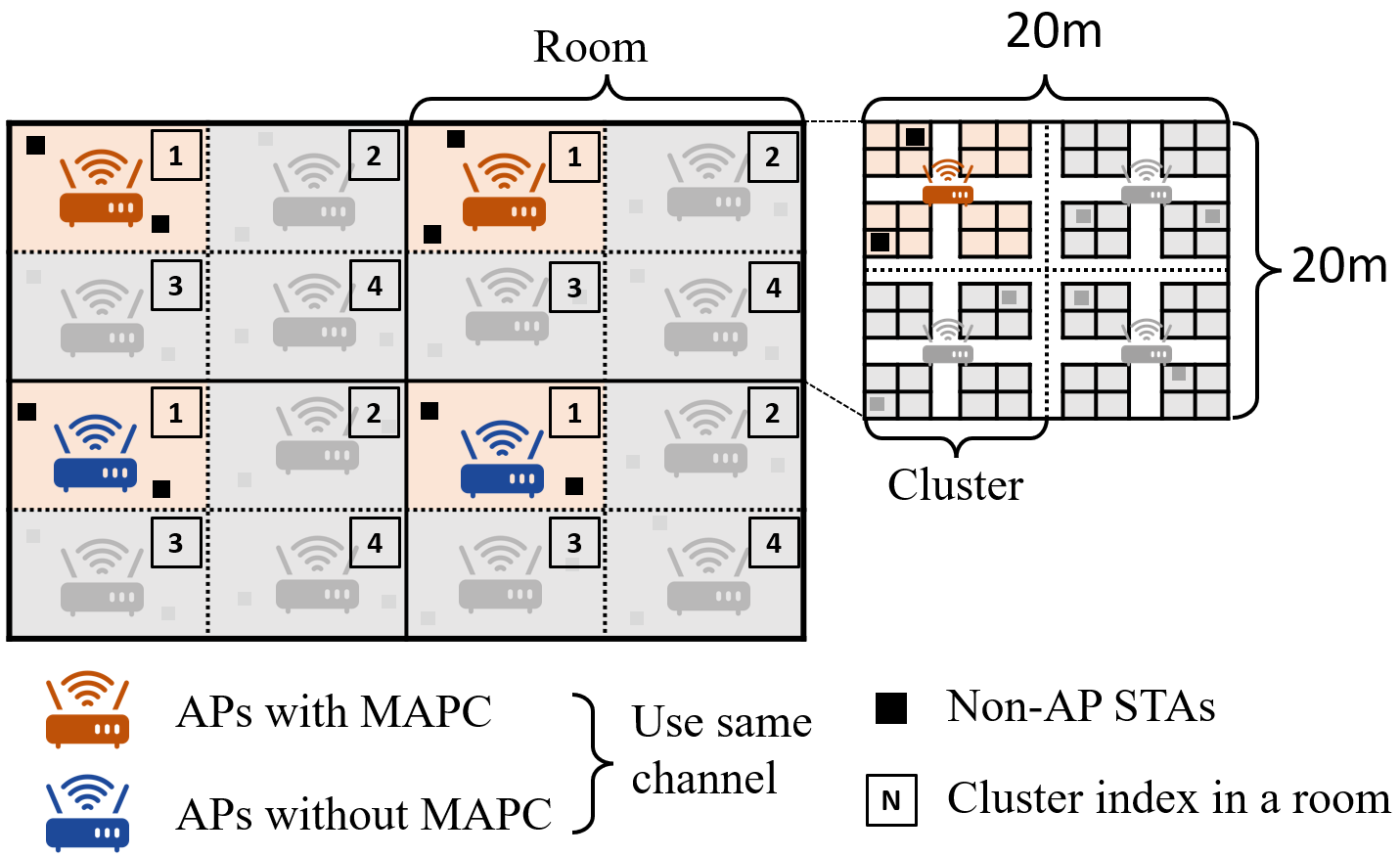}
    \caption{Topology (enterprise)}
    \label{fig:Topology}
    \vspace{-15pt}
\end{figure}

\vspace{-2pt}

\section{Simulation scenario}
\label{evaluation}
We investigate two comparable scenarios in an enterprise environment where collaboration takes place. In these scenarios, a single user demonstrates a real-time mobile gaming or VR application, while multiple users participate in an online meeting related to the demonstration. 

\begin{table}[t]
    \vspace{4pt}
    \centering
    \caption{Traffic parameters}
    \label{tab:trafficParams}
    \resizebox{0.8\columnwidth}{!}{
        \begin{tabular}{l|cccc|c}
            \toprule
            Traffic model & Direction & \makecell{AC} & \makecell{Avg. packet size \\(KBytes)} & \makecell{Avg. IAT \\(ms)} & Ref.\\
            \midrule
            RTMG &  DL & VO & 0.08 & 23.06 & ~\cite{Meng2019}\\
            RTMG &  UL & VO & 0.05 & 30.49 & ~\cite{Meng2019}\\
            VR & DL & VO & 166.66 & 33.33 & ~\cite{Lecci2021}\\
            VR & UL & VO & 0.19 & 10.81 & ~\cite{Lecci2021}\\
            VC & DL/UL & VI & 7.81 & 33.33 &~\cite{Porat2016}\\
            Background & DL/UL & BE & 200 & 8 &~\cite{Porat2016}\\
            \bottomrule
        \end{tabular}
    }
    \vspace{-7pt}
\end{table}

\begin{table}[t]
    \centering
    \caption{Simulations parameters}
    \label{tab:simParams}
    \scalebox{0.8}{
        \begin{tabularx}{\columnwidth}{l|Y}
            \toprule
            Parameter&Parameter Value\\
            \midrule
            Modulation order & \makecell{Data: MCS 7 \\ Ctrl: MCS 0} \\
            Transmitter power & 21 dBm \\
            Guard interval & 800 ns\\
            Frequency & 5 GHz\\
            Bandwidth & 80 MHz\\
            Max \# of A-MPDU & 64\\
            Max of PPDU duration & 5.484 ms\\
            Max \# of RUs & 4 (242-tone RUs) \\
            Simulation time & 5 s\\
            Iteration & 50 \\
            \bottomrule
        \end{tabularx}
    }
    \vspace{-15pt}
\end{table}

\subsection{Topology}
We adopt an enterprise environment used by TGax simulation scenarios for the practical topology~\cite{Merlin2015}. The environment that we consider is depicted in Fig.~\ref{fig:Topology}. An AP is positioned at the center of each cluster, with STAs randomly distributed within the cluster's area. BSSs located within the same cluster index in each room are assumed to operate on the same channel. We are especially interested in the BSSs within cluster index 1. Among these APs, two operate with MAPC, while the others do not.

\begin{figure*}[t]
    \centering
    \scalebox{0.88}{
    \begin{minipage}{0.329\linewidth}
        \centering
        \includegraphics[width=\linewidth]{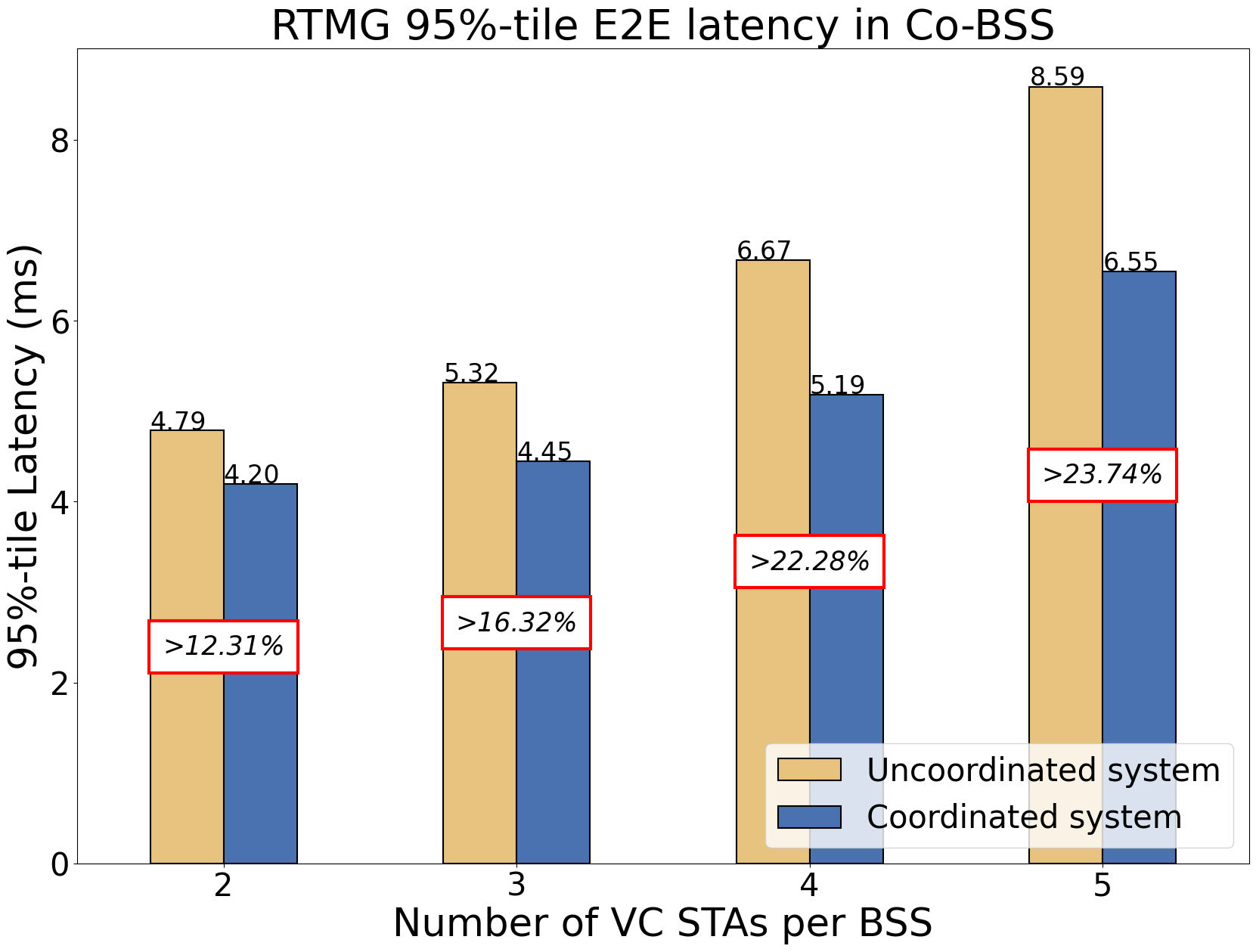}
        \subcaption{Co-BSS LL Latency}
        \label{fig:result-a}
    \end{minipage}
    \begin{minipage}{0.329\linewidth}
        \centering
        \includegraphics[width=\linewidth]{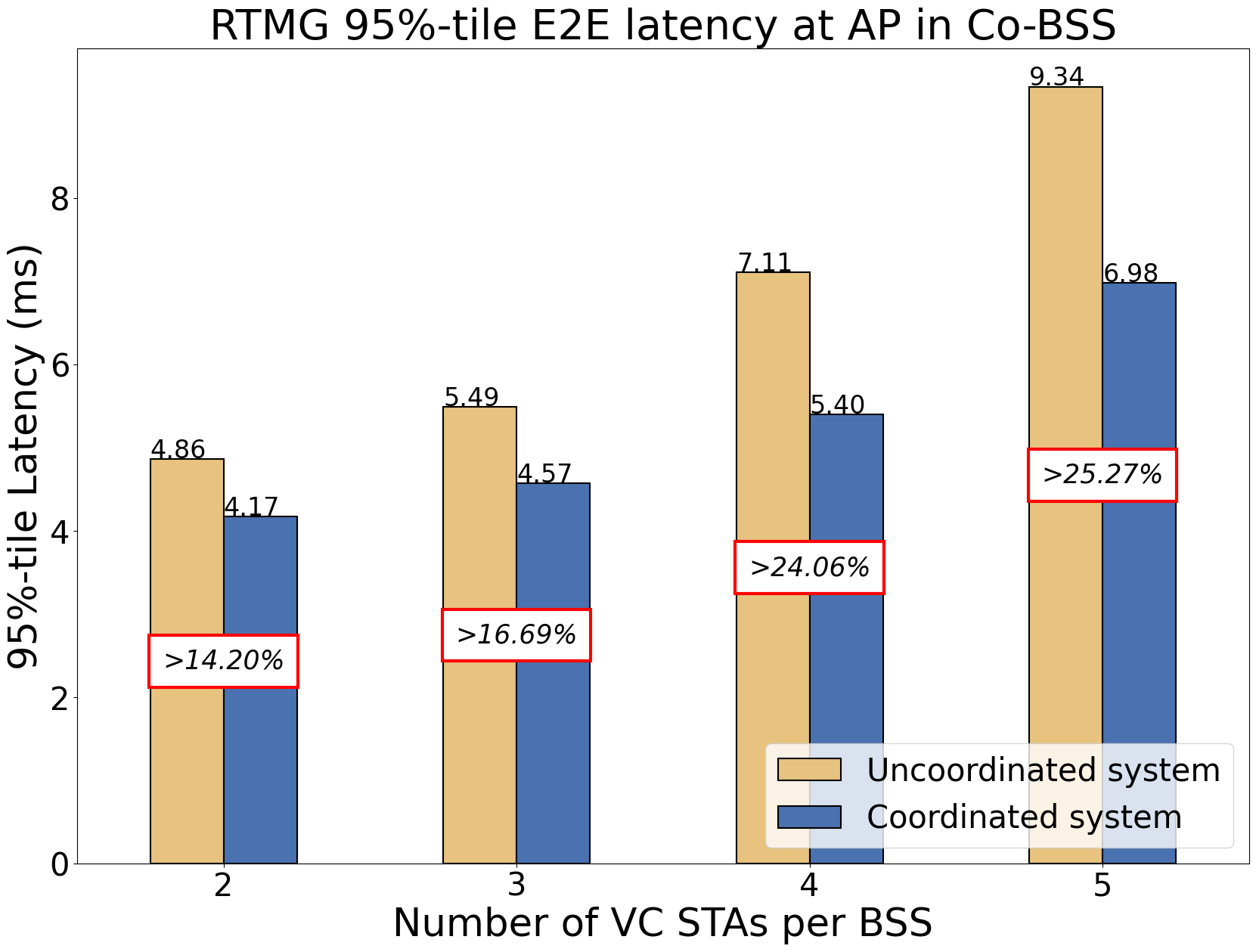}
        \subcaption{Co-BSS LL DL Latency}
        \label{fig:result-b}
    \end{minipage}
    \begin{minipage}{0.329\linewidth}
        \centering
        \includegraphics[width=\linewidth]{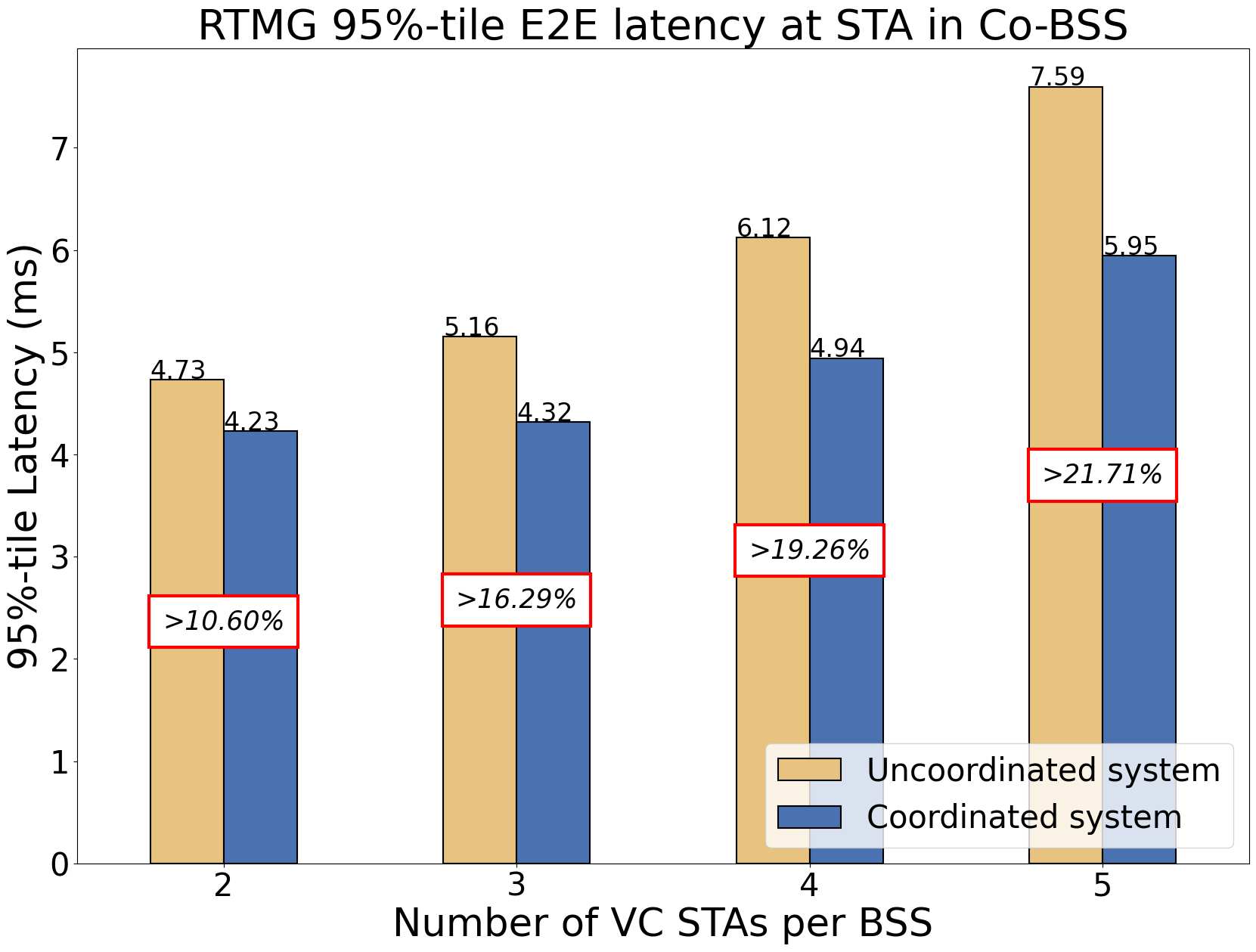}
        \subcaption{Co-BSS LL UL Latency}
        \label{fig:result-c}
    \end{minipage}
    }

    \vspace{0.15cm}

    \scalebox{0.88}{
    \begin{minipage}{0.329\linewidth}
        \centering
        \includegraphics[width=\linewidth]{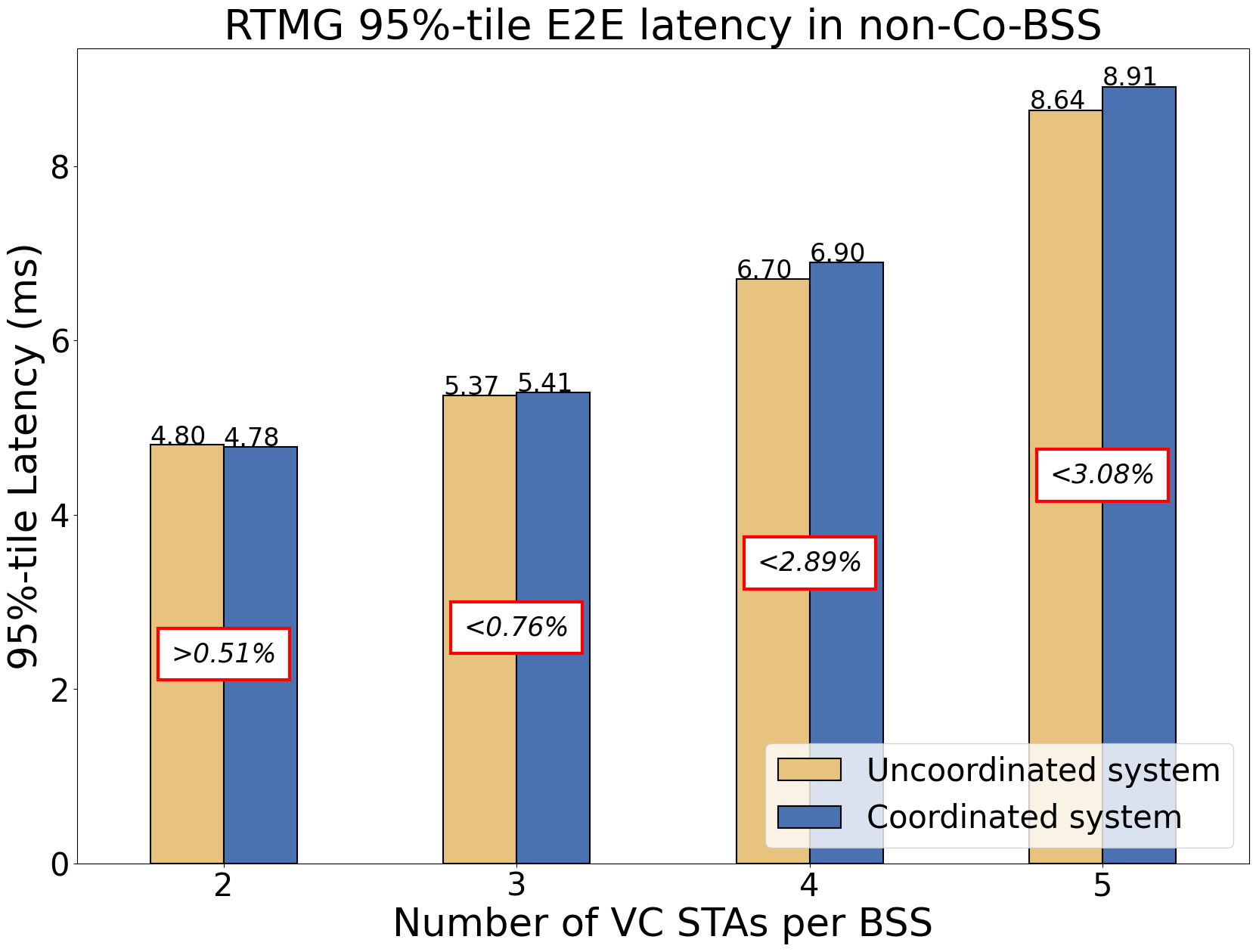}
        \subcaption{Non-Co-BSS LL Latency}
        \label{fig:result-d}
    \end{minipage}
    \begin{minipage}{0.329\linewidth}
        \centering
        \includegraphics[width=\linewidth]{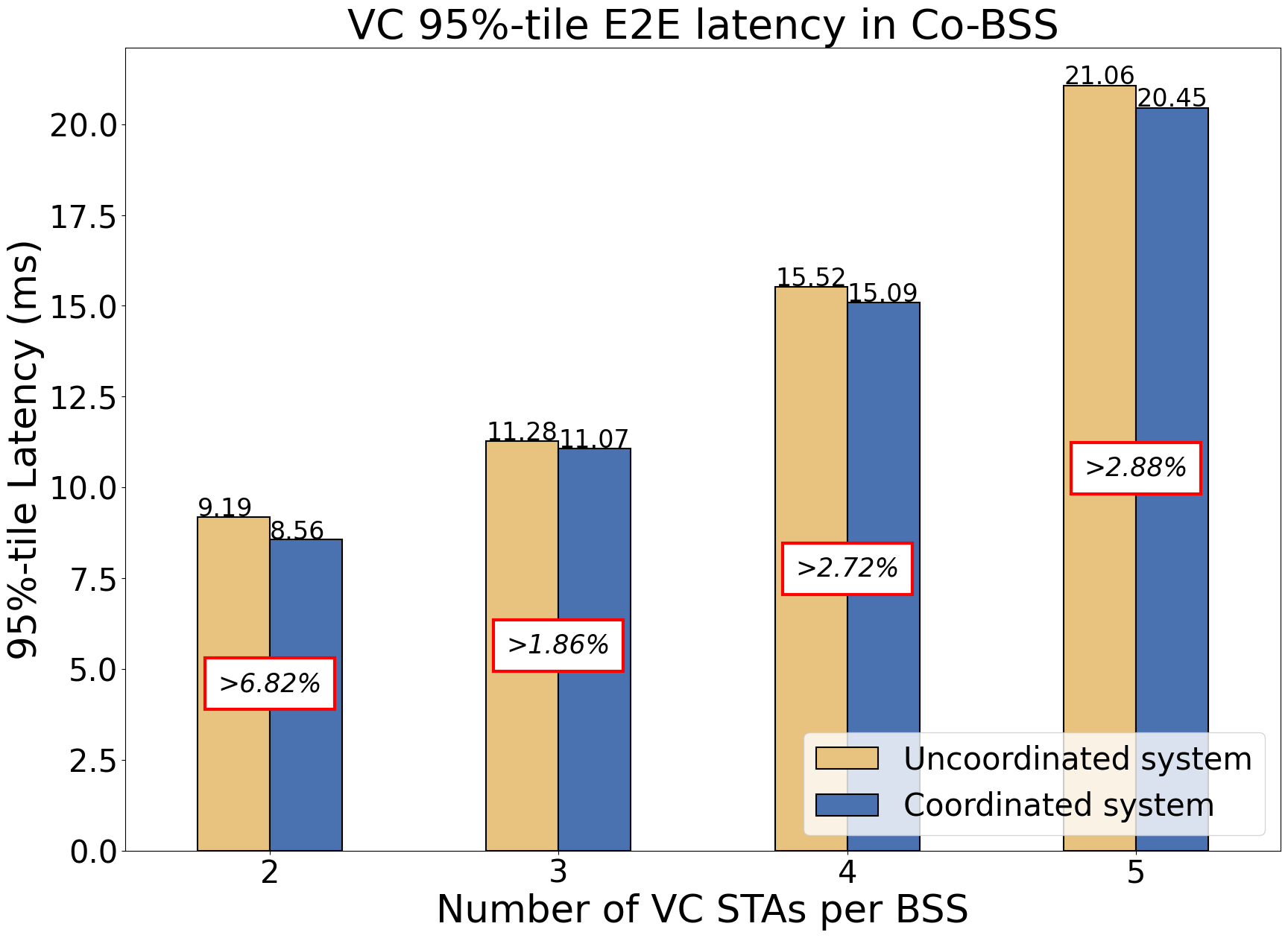}
        \subcaption{Co-BSS non-LL Latency}
        \label{fig:result-e}
    \end{minipage}
    \begin{minipage}{0.329\linewidth}
        \centering
        \includegraphics[width=\linewidth]{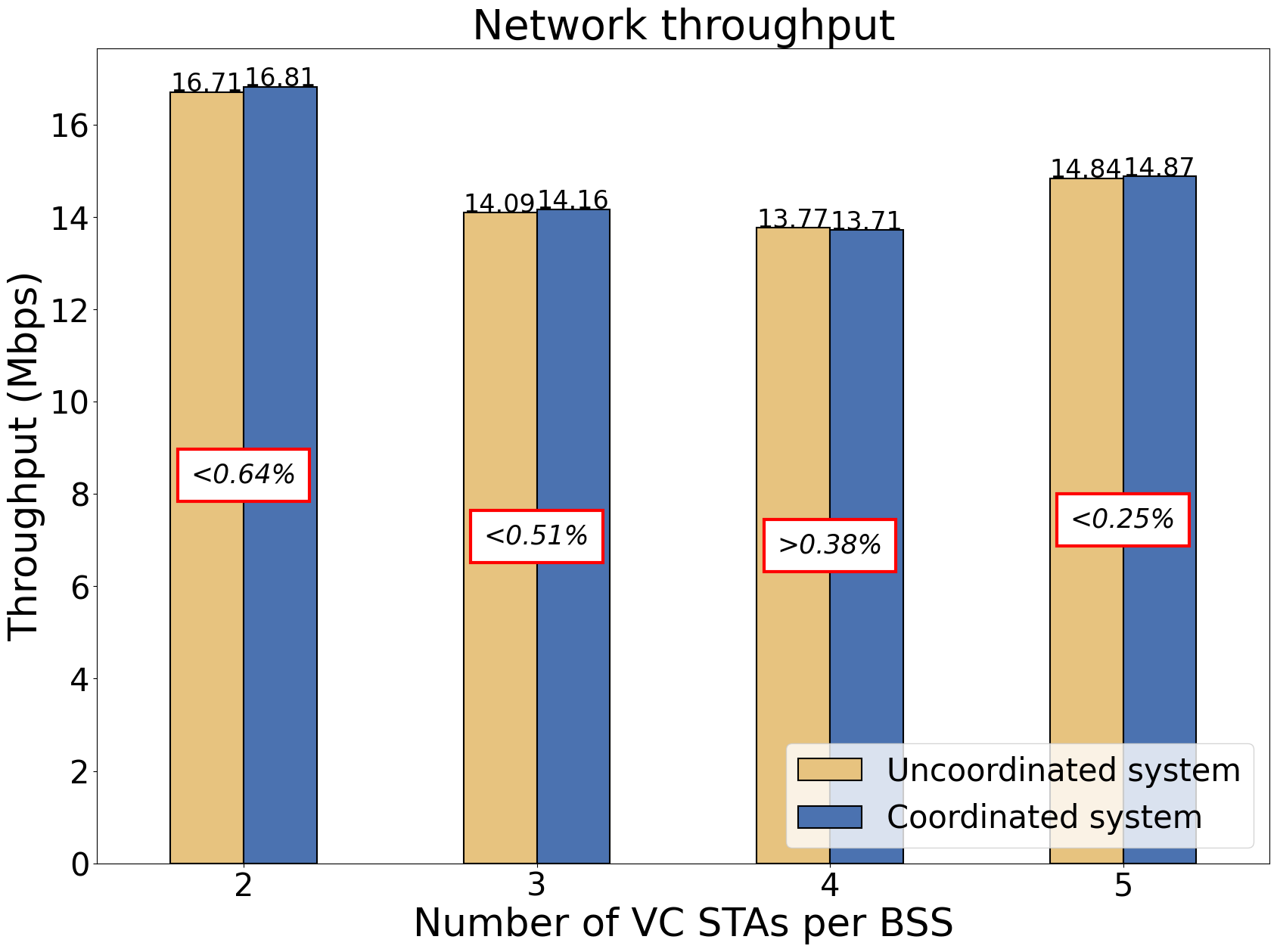}
        \subcaption{Network-wide throughput}
        \label{fig:result-f}
    \end{minipage}
    }
    \caption{Performance comparisons between the coordinated and uncoordinated systems according to the number of VC STAs per BSS in an RTMG traffic model scenario. (a)-(d) depict the 95\%-tile E2E latency experienced by RTMG traffic successfully transmitted in Co-BSS, by APs within Co-BSS, by STAs within Co-BSS, and in non-Co-BSS, respectively. (e) presents the 95\%-tile E2E latency of VC traffic successfully transmitted in Co-BSS. (f) shows the network-wide throughput.}
    \label{fig:result}
    \vspace{-15pt}
\end{figure*}

\subsection{Traffic models}
To analyze how Co-TDMA performance varies based on the characteristics of LL traffic generated by real-time applications, we consider two representative real-time application scenarios. First, we adopt the Status Sync Real-Time Mobile Gaming Traffic Model with a small traffic volume, referred to as RTMG traffic~\cite{Meng2019}. Second, we utilize the VR traffic with a large traffic volume derived from real-world data measurements as another real-time application scenario~\cite{Lecci2021}. Here, UL VR traffic was modeled periodically, following the approach used in~\cite{Alsakati2023}. Furthermore, we consider additional applications to reflect congested network conditions in each real-time application scenario. Video traffic is implemented based on the Video Conferencing Traffic Model defined in the TGax evaluation methodology~\cite{Porat2016}, referred to as VC traffic, while background traffic is generated using a constant bit rate (CBR) full-buffer model. The detailed parameters for traffic models are specified in Table~\ref{tab:trafficParams}. According to~\cite{Meng2019}, time-sensitive traffic is mapped to the voice (VO) AC. Thus, we assign RTMG and VR traffic to VO AC, while mapping VC traffic and background traffic to video (VI) AC and best effort (BE) AC, respectively. Packet size and inter-arrival time (IAT) follow their respective distribution models; however, only average values are presented in Table~\ref{tab:trafficParams}. Detailed distribution models are provided in~\cite{Meng2019, Lecci2021, Porat2016}.

\section{Results}

\subsection{Simulation methodology}

Here, we analyze the performance of E2E latency, and throughput in a coordinated system and an uncoordinated system. Particularly, we consider the 95th percentile (95\%-tile) E2E latency as an indicator of worst-case latency~\cite{Giordano2024, PAR, Naik2020}. Jitter is defined as the standard deviation of E2E latency~\cite{Meng2019}. Additionally, throughput is measured based on the MAC payload of successfully received packets at each STA, allowing us to assess the overall network throughput~\cite{Porat2016}.

We conducted system-level simulations to evaluate the performance of Co-TDMA under two real-time application scenarios: RTMG and VR scenarios. In each scenario, a coordinated system leveraging Co-TDMA and an uncoordinated system were simulated. 
We observed the performance gap between both systems in each real-time application scenario.
Each BSS consists of one STA for background traffic generation, N STAs for VC traffic transmission, and another STA responsible for RTMG or VR traffic, depending on the scenario.
This traffic distribution is identical across all BSSs. We refer to the STA with VC traffic as VC STA. To analyze the performance of the Co-TDMA under varying network congestion levels, we simulated by increasing the number of VC STAs per BSS from 2 to 5. All BSSs used the same TXOP limit set to its default parameter and used the UDP protocol as the transport protocol. Additional simulation parameters are summarized in Table~\ref{tab:simParams}.

\subsection{Performance analysis}

Fig.~\ref{fig:result} presents the simulation results for RTMG scenario in the network environment. Fig.~\ref{fig:result-a}-\ref{fig:result-c} show the 95\%-tile E2E latency experienced by RTMG traffic within coordinated BSSs (Co-BSSs), where MAPC is established. The X-axis represents the number of STAs providing VC traffic services per BSS, with a larger number indicating a higher level of network congestion.

The results indicate that Co-TDMA operation effectively reduces 95\%-tile E2E latency compared to the uncoordinated system. Furthermore, Fig.~\ref{fig:result-a} demonstrates that as network congestion increases, the gap in 95\%-tile E2E latency between the coordinated system and the uncoordinated system increases.
This indicates that Co-TDMA becomes increasingly effective as channel access contention intensifies. However, the growth rate of the gap gradually diminishes since ${T}^{\text{Busy}}_{s,k}$ increases logarithmically with network congestion, as discussed in Section~\ref{co-tdma scheme}.

The overall latency reduction can be further analyzed by separating DL and UL latency variations, as shown in Fig.~\ref{fig:result-b} and Fig.~\ref{fig:result-c}. 
It shows that DL LL traffic experiences greater performance gains than UL LL traffic in the Co-BSS. However, it is noteworthy that the worst-case latency of UL traffic is also substantially improved.
It implies that Co-TDMA can play a crucial role in supporting UL traffic efficiently in real-time applications\cite{Evgeny2019}.


Fig.~\ref{fig:result-d} and Fig.~\ref{fig:result-e} illustrate the latency variation for traffic that is indirectly affected by Co-TDMA. As shown in Fig.~\ref{fig:result-d}, LL traffic handled in BSSs without MAPC experienced slightly higher latency in the coordinated system. In contrast, Fig.~\ref{fig:result-e} shows that non-LL traffic (i.e., VC traffic) transmitted in Co-BSSs exhibits reduced latency. These results imply that Co-TDMA allows APs in Co-BSSs to use the network time resource more efficiently.

The overall network throughput, considering all four BSSs, is presented in Fig.~\ref{fig:result-f}. The results indicate that Co-TDMA operation does not significantly impact total network throughput. Ultimately, simulation results validate that Co-TDMA reduces the latency of RTMG traffic with small-volume LL characteristics by approximately 12–24\%, while minimally affecting overall network performance.

\begin{table}[t]
    \vspace{4pt}
    \centering
    \caption{95\%-tile E2E latency (ms) in Co-BSSs}
    \label{tab:traffi_model_differ}
    \resizebox{0.88\columnwidth}{!}{
        \begin{tabular}{l|c|cccc}
            \toprule
            Scenario & System & 2 VC STAs & 3 VC STAs & 4 VC STAs & 5 VC STAs \\
            \midrule
            RTMG &  \makecell{Uncoordinated \\ Coordinated}
            & \makecell{4.79 \\ 4.20}  
            & \makecell{5.32 \\ 4.45}  
            & \makecell{6.67 \\ 5.19}  
            & \makecell{8.59 \\ 6.55} \\ 
            \midrule
            VR &  \makecell{Uncoordinated \\ Coordinated}   
            & \makecell{17.82 \\ 16.60}  
            & \makecell{22.95 \\ 21.13} 
            & \makecell{28.83 \\ 27.57} 
            & \makecell{39.08 \\ 35.19} \\ 
            \bottomrule
        \end{tabular}
    }
    \vspace{-13pt}
\end{table}

\begin{table}[t]
    \vspace{4pt}
    \centering
    \caption{Jitter (ms) in Co-BSSs}
    \label{tab:traffi_model_differ_2}
    \resizebox{0.88\columnwidth}{!}{
        \begin{tabular}{l|c|cccc}
            \toprule
            Scenario & System & 2 VC STAs & 3 VC STAs & 4 VC STAs & 5 VC STAs \\
            \midrule
            RTMG &  \makecell{Uncoordinated \\ Coordinated}
            & \makecell{1.99 \\ 1.51}  
            & \makecell{2.42 \\ 1.68}  
            & \makecell{3.88 \\ 2.27}  
            & \makecell{6.38 \\ 3.63} \\ 
            \midrule
            VR &  \makecell{Uncoordinated \\ Coordinated}   
            & \makecell{24.43 \\ 21.10}  
            & \makecell{41.12 \\ 34.70} 
            & \makecell{66.77 \\ 60.68} 
            & \makecell{126.87 \\ 107.59} \\ 
            \bottomrule
        \end{tabular}
    }
    \vspace{-18pt}
\end{table}

Table~\ref{tab:traffi_model_differ} presents the 95\%-tile E2E latency experienced by LL traffic in Co-BSSs under both the RTMG and VR scenarios. While the overall trend in the VR scenario is similar to that of the RTMG scenario, the relative reduction ratio between the coordinated and uncoordinated systems is less pronounced in the VR scenario.

Table~\ref{tab:traffi_model_differ_2} illustrates the jitter experienced by LL traffic in Co-BSSs for both scenarios. A smaller jitter value means limited variation in latency, implying more predictable communication. In both scenarios, the coordinated system shows reduced jitter, indicating that Co-TDMA contributes to more stable communication required for real-time applications~\cite{Meng2019,Giordano2024}. In particular, a greater reduction ratio in jitter is observed in the RTMG scenario compared to the VR scenario.

We anticipate that the volume of LL traffic contributes to the observed results. As shown in Table~\ref{tab:trafficParams}, VR traffic has a significantly larger volume than RTMG traffic. Consequently, VR traffic is more likely to exceed the available shared TXOP duration, making it difficult to service buffered traffic fully within the duration. That is, VR traffic cannot fully benefit from the channel access delay gain analyzed in Section~\ref{co-tdma scheme}, which leads to the smaller performance improvement compared to the RTMG scenario.
\vspace{-5pt}

\section{Conclusion}
\vspace{-3pt}
In this paper, we investigated Co-TDMA operation, a key mechanism introduced in IEEE 802.11bn (Wi-Fi 8) for next-generation wireless networks. Our results clearly demonstrate that Co-TDMA is not only essential but also highly effective in reducing the latency of LL traffic under congested network conditions. In particular, the 95\%-tile latency and jitter of LL traffic are significantly reduced, thereby better supporting real-time applications. We revealed that Co-TDMA performance is affected by network conditions, particularly congestion from multiple STAs and LL traffic characteristics. To validate these effects, we conducted system-level simulations under an enterprise collaboration scenario.

The proposed scheduling method can be utilized as a foundation for developing scheduling strategies that satisfy the specific requirements of real-time applications, such as delay bounds. As future work, we plan to scale the current two-MAPCs scenario to MAPC scenarios with more APs,
analyze the requirements of individual real-time applications, and develop scheduling algorithms tailored to meet those demands.
\vspace{-5pt}

\bibliographystyle{IEEEtran}
\bibliography{Globecom-2025-ref}

\vspace{12pt}

\end{document}